\documentclass[doublecol,preprintnumbers,nofootinbib]{epl2} 
\usepackage{amssymb}
\usepackage{amsmath}
\usepackage{graphicx}
\usepackage{subfigure}
\usepackage{color}
\title{Higgs to \mbox{\boldmath $\mu^\mp \tau^\pm$} Decay in Supersymmetry without \mbox{\boldmath $R$} Parity}
\author{Abdesslam Arhrib\inst{1} \and Yifan Cheng\inst{2} \and Otto C. W. Kong\inst{2}}
\shortauthor{A. Arhrib, Y. Cheng and O. C. W. Kong}
\institute{                    
  \inst{1} Department of Mathematics, Faculty of Science and Techniques, B.P
  416 Tangier, Morocco. \\
  \inst{2} Department of Physics and 
Center for Mathematics and Theoretical Physics,
National Central University, Chung-Li, Taiwan 32054.}
\pacs{14.80.Da}{Supersymmetric Higgs bosons}
\pacs{11.30.Hv}{Flavor symmetries}
\pacs{12.60.Jv}{Supersymmetric models}
\abstract{
In this letter, we report on lepton flavor violating Higgs decay into
 $\mu^\mp \tau^\pm$ in the framework of the generic supersymmetric 
standard model without R parity and list {interesting} combinations of 
R-parity violating parameters. We impose other known experimental
constraints on the parameters of the model and show
our results from the R-parity 
violating parameters. In our analysis, the branching ratio of 
$h^0 \rightarrow \mu^\mp \tau^\pm$ can exceed $10^{-5}$ 
within admissible parameter space.}

\begin{document}
\maketitle

\section{Introduction}
In the Standard Model (SM), the lepton number of each flavor is separately conserved. However, it is 
well known from the neutrino oscillation experiments that lepton flavor 
conservation should be violated \cite{superk}, which also gives hints of physics beyond the 
SM. Therefore, various lepton flavor violating (LFV) processes have received much attention. For example, in the framework of supersymmetry, which
is undoubtedly the most popular candidate theory for physics beyond SM, flavor 
violating processes such as $\tau\rightarrow\mu\gamma$, 
$\tau\rightarrow\mu X$, $\tau\rightarrow\mu\eta$, etc. are 
discussed in the minimal supersymmetric standard model (MSSM) \cite{other} and other 
supersymmetric extensions \cite{otherright,othermodel}. 
Specifically, LFV Higgs 
to $\mu^\mp \tau^\pm$ decay which is the focus of 
this letter, has been studied using the effective Lagrangian method in Ref.\cite{Blankenburg};
the general two Higgs doublet model \cite{tao,2hdm} which allows lepton flavor violation; 
MSSM \cite{2hdm,Brignole}; and MSSM with right-handed neutrinos \cite{seesaw}.
Though supersymmetry so far lacks experimental evidence \cite{susyex} and 
hence faces stringent challenges, it is pointed out \cite{alive} that there is 
still room for the (minimal) supersymmetric standard model to accommodate existing 
experimental results. For example, large mass spectrum around or beyond 1 TeV for a
majority of supersymmetric particles is possible. Allowing non-universal soft supersymmetry
breaking masses or more free parameters can also relax the constraints imposed by experiments.

In many studies of the supersymmetric standard model, a discrete symmetry called ``R parity", 
which keeps baryon and lepton number conservation, is often imposed to prevent proton 
decay and make the lightest supersymmetric particle (LSP) a dark matter candidate. From 
a theoretical point of view, R parity is {\it ad hoc}, not motivated so long
as the phenomenological (minimal) supersymmetric standard model is concerned \cite{Perez}.
Otherwise, a generic version of (minimal) supersymmetric standard model without R parity, 
which is our focus here, has the advantage of a richer phenomenology and enabling neutrino masses 
and mixings without the need to introduce any extra superfield. Beyond neutrino masses, the 
key part of R-parity violation of interest is lepton flavor violation, because it can be 
observed in processes involving only SM particles 
while SM itself does not give rise to these kinds of 
processes. Higgs to $\mu^\mp \tau^\pm$ decay is our example at
hand. In the framework of R-parity violating (RPV) supersymmetry, 
there have been some studies \cite{rparity, ottorparity, ottoleptonic} 
on the issue of lepton flavor violating processes. 
However, such studies either were limited to particular types of R-parity 
violation or did not study Higgs to $\mu^\mp \tau^\pm$. 
Recently both ATLAS and CMS reported discovery of a boson state with mass 
near 125-126 GeV \cite{ATLAS,CMS}.
The observed boson is essentially compatible 
with a SM-like Higgs, 
while more data is needed to fully pin down its nature.
The flavor violating decay $h^0\rightarrow \mu^\mp \tau^\pm$ 
is of particular interest at this moment.

In this short letter, we present the first results
of a comprehensive study on the generic supersymmetric standard model, without R parity, 
highlighting cases of most interest. Details of our studies will be reported in another
publication \cite{next}.

In the following section, we summarize the basic formulation and parametrization of the generic 
supersymmetric standard model, without R parity. The neutral scalar mass matrix is particularly discussed, 
including the key quantum corrections up to two-loop order. We then sketch our calculations
and present our numerical results in section 3. Note that we made no 
assumptions on RPV parameters, hence our analysis did not lose general validity while the Higgs mass and 
the mass spectrum of the other particles are kept within experimental constraints. Finally, we 
summarize this letter and give an account of upcoming works.

\section{Generic Supersymmetric Standard Model (without R Parity) and Higgs Mass Matrix}
With the superfield content of the MSSM,
the most general renormalizable superpotential can be written as
\begin{align}
W=&\,\epsilon_{ab}
\left[\right.
\mu_\alpha \hat{H}^a_u \hat{L}^b_\alpha + h^u_{ik} \hat{Q}^a_i \hat{H}^b_u \hat{U}^C_k
+ \lambda^{'}_{\alpha jk} \hat{L}^a_\alpha \hat{Q}^b_j \hat{D}^C_k  \nonumber\\
&+ \frac{1}{2}\lambda_{\alpha\beta k}\hat{L}^a_\alpha \hat{L}^b_\beta \hat{E}^C_k
\left.\right] 
+ \frac{1}{2}\lambda^{''}_{ijk}\hat{U}^C_i \hat{D}^C_j \hat{D}^C_k 
\end{align}
where $(a,b)$ are SU(2) indices with $\epsilon_{12}=- \epsilon_{21}=1$, $(i,j,k)$ are the 
usual family (flavor) indices, and $(\alpha ,\beta)$ are extended flavor indices ranging from 
0 to 3. Note that $\lambda$ is antisymmetric in the first two indices as required by SU(2) 
product rules, while $\lambda^{''}$ is antisymmetric in the last two indices by $\text{SU(3)}_C$\,.
The soft supersymmetry breaking terms can be written as follows: 
\begin{align}
V\!&_{\rm soft} =\nonumber\\
&\Big\{ \epsilon_{ab}
  B_{\alpha} \,  H_{u}^a \tilde{L}_\alpha^b \nonumber\\
  &\;+\epsilon_{ab} \left[ \,
  A^U_{ij} \tilde{Q}^a_i H^b_u \tilde{U}^\dagger_j
 +A^D_{ij} H^a_d \tilde{Q}^b_i \tilde{D}^\dagger_j 
 +A^E_{ij} H^a_d \tilde{L}^b_i \tilde{E}^\dagger_j \right] \nonumber\\
&\;+\epsilon_{ab}\left[  
  A^{\lambda '}_{ijk} \tilde{L}^a_i \tilde{Q}^b_j \tilde{D}^\dagger_k 
 +\frac{1}{2} A^\lambda_{ijk} \tilde{L}^a_i \tilde{L}^b_j \tilde{E}^\dagger_k \right]
 \!+\!\frac{1}{2} A^{\lambda ''}_{ijk} \tilde{U}^\dagger_i \tilde{D}^\dagger_j \tilde{D}^\dagger_k\nonumber \\
&\; + \frac{M_1}{2} \tilde{B}\tilde{B}
   + \frac{M_2}{2} \tilde{W}\tilde{W}
   + \frac{M_3}{2} \tilde{g}\tilde{g}
+ \text{h.c.} \big.\Big\} \nonumber\\
&+
 \tilde{Q}^\dagger \tilde{m}_Q^2 \,\tilde{Q} 
+\tilde{U}^{\dagger} 
\tilde{m}_U^2 \, \tilde{U} 
+\tilde{D}^{\dagger} \tilde{m}_D^2 
\, \tilde{D} \nonumber\\
&+ \tilde{L}^\dagger \tilde{m}_L^2  \tilde{L}  
  +\tilde{E}^{\dagger} \tilde{m}_E^2 
\, \tilde{E}
+ \tilde{m}_{H_u}^2 \,
|H_{u}|^2 
\; ,
\end{align}
where $\tilde{m}^2_L$ is given by a 4 $\times$ 4 
matrix with zeroth components.  $\tilde{m}^2_{L_{00}}$ corresponds to 
$\tilde{m}^2_{H_d}$ in MSSM, while $\tilde{m}^2_{L_{0k}}$'s give new mass mixings. {Note that $\tilde{U}^\dagger$, $\tilde{D}^\dagger$, and $\tilde{E}^\dagger$ are the scalar components of the superfields $\hat{U}^C$, $\hat{D}^C$, and $\hat{E}^C$, respectively.}

The above, together with the standard (gauged) kinetic terms, describe the full Lagrangian of the
model. We have four leptonic superfields $\hat{L}$, which contain the components of fermion doublet 
as $l^0$ and $l^-$, and their scalar partners as $\tilde{l}^0$ and $\tilde{l}^-$. For convenience, 
we choose a flavor basis such that only $\hat{L}_0$ bears a nonzero vacuum expectation value (VEV) 
and thus can be identified as the usual $\hat{H}_d$ in the MSSM. 
However, one should keep in mind that 
$\hat{H}_d$ now may contain the partly charged lepton states. 
The parametrization has the advantage that tree level RPV contributions to {neutral scalar} mass matrix are 
described completely by the $\mu_i$, $B_i$, and $\tilde{m}^2_{L_{0i}}$ parameters, 
which are well constrained to be small even with very conservative neutrino mass bounds 
imposed \cite{Otto,otto98}. The tree level mass matrices for the {charged} scalar fields also have minimal
extra terms from R-parity violation \cite{Otto}.

For the mass matrices for the (colorless) scalars, we have five 
neutral complex fields, from $\hat{H_u}$ and four 
$\hat{L}_\alpha$ superfields, and eight charged fields. Explicitly, for the 
neutral part we write the  $(1+4)$ complex fields 
in terms of their scalar and pseudoscalar parts, in the order 
($h^{0\dagger}_u$, $\tilde{l}^0_0$, $\tilde{l}^0_1$, 
$\tilde{l}^0_2$, $\tilde{l}^0_3$), 
to form a full $10 \times 10$ (real and symmetric) 
mass-squared matrix. As for charged scalars, the basis 
$\{ h^{+\dagger}_u, \tilde{l}^-_0, \tilde{l}^-_1, 
\tilde{l}^-_2, \tilde{l}^-_3, \tilde{l}^{+\dagger}_1, 
\tilde{l}^{+\dagger}_2, \tilde{l}^{+\dagger}_3 \}$ 
is used to write the $8 \times 8$ mass-squared matrix. 
The exact form of these tree level 
matrix elements can be found in \cite{Otto}.

With all the RPV terms, the physical scalar states 
are now a mixture of Higgses and sleptons. 
The RPV terms provide new contributions to scalar mass matrices and 
hence Higgs mass. In addition, third generation quarks and squarks 
could play an important role in 
radiative corrections to the Higgs sector, and hence should be included.
Accordingly, in our computation, 
we implement complete one-loop corrections, 
from Ref.\cite{Ellis}, for CP-even, CP-odd and charged Higgs bosons.
Moreover, the light Higgs mass 
issue is quite sensitive and needs to be treated delicately, 
therefore we include an estimation \cite{2loop}
of the key two-loop corrections for the light Higgs boson
\footnote{The Higgs bosons mix with sleptons via RPV terms, but we can still identify 
the states from other sleptons due to the foreseeable smallness of RPV parameters.}. 
Note that radiative RPV corrections are typically too small to 
be taken into account, 
thus we study tree level RPV effects only.

\section{Calculations and Numerical results}
In this section, we describe the most interesting RPV parameter 
combinations, {and the numerical results of their contribution to $h^0\to\mu^\mp \tau^\pm$. Under the framework of R-parity violating supersymmetry, we can have the LFV Higgs decays at tree level, while the effective coupling between a neutral scalar and two charged fermions/leptons can be written as
\begin{equation}
{\cal L}=g_{\scriptscriptstyle 2}\overline{\Psi}(\chi^-_{\bar{n}})
 \left[{\cal C}_L\frac{1-\gamma_5}{2}
      +{\cal C}_R\frac{1+\gamma_5}{2} \right] \Psi(\chi^-_n)\Phi(\phi^0_m) 
\end{equation}      
where
\begin{align}
{\cal C}_L=
&-\frac{1}{\sqrt{2}}\mbox{\boldmath $V$}^*_{2\bar{n}}   
        \mbox{\boldmath $U$}_{\!1n}({\cal D}^s_{1m}+i{\cal D}^s_{6m}) \nonumber\\
&-\frac{1}{\sqrt{2}}\mbox{\boldmath $V$}^*_{1\bar{n}}\mbox{\boldmath $U$}_{\!2n}
        ({\cal D}^s_{2m}-i{\cal D}^s_{7m})   \nonumber\\
&-\frac{1}{\sqrt{2}}\mbox{\boldmath $V$}^*_{1\bar{n}}\mbox{\boldmath $U$}_{\!\left(j+2\right)n} 
        ({\cal D}^s_{\left(j+2\right)m}-i{\cal D}^s_{\left(j+7\right)m})    \nonumber\\
&-\frac{y_{e_j}}{\sqrt{2}g_{\scriptscriptstyle 2}}\mbox{\boldmath $V$}^*_{\left(j+2\right)\bar{n}}
        \mbox{\boldmath $U$}_{\!\left(j+2\right)n}({\cal D}^s_{2m}+i{\cal D}^s_{7m}) \nonumber\\       
&+\frac{y_{e_j}}{\sqrt{2}g_{\scriptscriptstyle 2}}\mbox{\boldmath $V$}^*_{\left(j+2\right)\bar{n}}
        \mbox{\boldmath $U$}_{\!2n} ({\cal D}^s_{\left(j+2\right)m}+i{\cal D}^s_{\left(j+7\right)m}) \nonumber
\end{align}
\begin{align}         
&+\frac{\lambda_{ijk}}{\sqrt{2}g_{\scriptscriptstyle 2}}\mbox{\boldmath $V$}^*_{\left(k+2\right)\bar{n}}
        \mbox{\boldmath $U$}_{\!\left(i+2\right)n}({\cal D}^s_{\left(j+2\right)m}+i{\cal D}^s_{\left(j+7\right)m}) 
\nonumber\\
{\cal C}_R=
&-\frac{1}{\sqrt{2}}\mbox{\boldmath $U$}^*_{1\bar{n}} \mbox{\boldmath $V$}_{\!2n}
        ({\cal D}^s_{1m}-i{\cal D}^s_{6m}) \nonumber\\
&-\frac{1}{\sqrt{2}}\mbox{\boldmath $U$}^*_{2\bar{n}}\mbox{\boldmath $V$}_{\!1n}   
        ({\cal D}^s_{2m}+i{\cal D}^s_{7m}) \nonumber\\
&-\frac{1}{\sqrt{2}}\mbox{\boldmath $U$}^*_{\left(j+2\right)\bar{n}} \mbox{\boldmath $V$}_{\!1n}  
        ({\cal D}^s_{\left(j+2\right)m}+i{\cal D}^s_{\left(j+7\right)m}) \nonumber\\
&-\frac{y_{e_j}}{\sqrt{2}g_{\scriptscriptstyle 2}} \mbox{\boldmath $U$}^*_{\left(j+2\right)\bar{n}}
        \mbox{\boldmath $V$}_{\!\left(j+2\right)n}({\cal D}^s_{2m}-i{\cal D}^s_{7m}) \nonumber\\
&+\frac{y_{e_j}}{\sqrt{2}g_{\scriptscriptstyle 2}}\mbox{\boldmath $U$}^*_{2\bar{n}} 
        \mbox{\boldmath $V$}_{\!\left(j+2\right)n}({\cal D}^s_{\left(j+2\right)m}-i{\cal D}^s_{\left(j+7\right)m})
        \nonumber\\ 
&+\frac{\lambda^*_{ijk}}{\sqrt{2}g_{\scriptscriptstyle 2}} \mbox{\boldmath $U$}^*_{\left(i+2\right)\bar{n}} 
        \mbox{\boldmath $V$}_{\!\left(k+2\right)n}
        ({\cal D}^s_{\left(j+2\right)m}-i{\cal D}^s_{\left(j+7\right)m}) \,.
\end{align}      
${\cal D}^{s}$ diagonalizes the neutral scalar mass matrix ${\cal M}^{2}_{S}$ as ${{\cal D}^{s}}^{T}{\cal M}^{2}_{S}{\cal D}^{s}$ = diag\{${M^2_S}\}$. $m$ ranges from 1 to 10, indicating the 10 neutral scalar mass eigenstates (including the light Higgs and a goldstone boson state). \mbox{\boldmath $V$} and \mbox{\boldmath $U$} are two unitary matrices which diagonalize the 5 $\times$ 5 charged fermion matrix ${\cal M}_C$ as $\mbox{\boldmath $V$}^\dagger{\cal M}_C\mbox{\boldmath $U$}=\text{diag}\left\{M_{\chi^-_n} \right\} \equiv \text{diag} \left\{M_{c1},M_{c2},m_e,m_\mu,m_\tau \right\}$. $n$ and $\bar{n}$ range from 1 to 5, corresponding to the mass eigenstates of 2 charginos and 3 charged leptons \cite{Otto}.}

{There are three types of combinations, namely $B_i \mu_j$, $B_i\lambda$ and $\mu_i \mu_j$, which can contribute to $h^0\to\mu^\mp \tau^\pm$ at tree level already. In addition to the three combinations above, $B_i A^\lambda$ also provides a possible approach to $h^0\to\mu^\mp \tau^\pm$ which may be quite interesting since $A^\lambda$ does not have known experimental constraints. In this letter, we highlight these four cases and study their contributions to the LFV Higgs to $\mu^\mp \tau^\pm$ decay up to one-loop level.}

In our numerical computation, we deal directly with mass eigenstates and put all the
tree level mass matrices into the program. In the case of Higgs masses, we include as
well the one-loop and two-loop corrections to relevant mass matrix elements as mentioned in
the second section. The mass of the Higgs bosons (and other sparticles) needed in our analysis 
are obtained by diagonalizing corresponding mass matrices numerically.
The necessary amplitudes of tree and one-loop Feynman diagrams \footnote{During numerical computation of Feynman diagrams, \textit{LoopTools} package 
is used for the evaluation of loop integrals \cite{looptools}.}
and relevant effective couplings in the model are 
derived analytically.
By encoding the derived analytical formulas of the
decay amplitudes into the numerical program, values of total amplitude and
hence decay rate can be obtained. In the computation of the total decay width
of light Higgs, we include all significant decay channel such as $\bar{b}b$,
$\tau^-\tau^+$, $WW^*$, $ZZ^*$, $\gamma\gamma$, and $gg$,  
as well as the RPV decay $h^0\rightarrow \mu^\mp \tau^\pm$.
With the RPV partial decay width rate for the channel and total 
decay width, the branching ratio can be obtained.

There are many different sources of experimental constraints 
on the RPV parameter space. We have taken into consideration all the available experimental constraints, e.g., flavor violating charged lepton decays like $\tau^-\to\mu^-e^+e^-$ \cite{ottomu} and $\tau\to\mu\gamma$ \cite{ottoleptonic}, 
experimental values of CKM matrix elements and various other proocesses \cite{Barbier} to confine the 
admissible RPV parameter values. Especially, the constraint from indirect evidence of neutrino 
mass, i.e., $\sum_i m_{\nu_i}\lesssim\text{1 eV}$ 
\cite{neutrino} is quite crucial. If we do not consider the neutrino mass constraint, the constraints from rare tau decays are not stringent and hence can admit a large branching ratio of the LFV Higgs decay, as indicated by Ref. \cite{Harnik}. Meanwhile, as discussed before, ATLAS and CMS data \cite{ATLAS,CMS} are pointing
out a Higgs-like boson with a mass around 125-126 GeV. After including 
the uncertainties on the experimental Higgs mass, and loop corrections 
estimation to Higgs mass matrix, we allow the light Higgs 
mass to be in the range of 123 to 127 GeV. 
Other restrictions and assumptions we used are as follows:
$\left|\mu_0\right|$, $\left|A_u\right|$, $\left|A_d\right|$ and 
$\left|A^\lambda\right|\leq\text{2500 GeV}$; $A_e$ is set to be zero since 
its influence is quite small and negligible; $\tilde{m}^2_E=\tilde{m}^2_L$ 
(without zeroth component) $\leq(\text{2500 GeV})^2$ with off-diagonal 
elements zero; $\tan\beta$ ranges from 3 to 60; heavy Higgs/sneutrino 
and charged Higgs/slepton masses are demanded to be above 200 GeV.
We use the 
relation $M_2=\frac{1}{3.5}M_3=2M_1$ between 
three gaugino masses and the condition that squarks of the first two families 
cannot be lighter than about $0.8 M_3$. Therefore we take soft 
supersymmetry breaking scalar masses 
$\tilde{m}^2_Q=\tilde{m}^2_U=\tilde{m}^2_D=(0.8 M_3\times \text{identity matrix})^2$ for simplicity.
The parameter setting is in accordance with the
gravity-mediated supersymmetry breaking picture \cite{Martin}, for instance.
We are interested in using a concrete setting that is compatible with known
constraints but otherwise not too restrictive, to illustrate what we expect to be
more generic features of the RPV signature. 

We discuss below the most interesting combinations of RPV
parameters and the numerical results we obtained.

\subsection{Contributions from $B_i\,\mu_j$ combinations}
\begin{figure}[ht]
\begin{minipage}[]{0.327\linewidth} 
\includegraphics[scale=0.51]{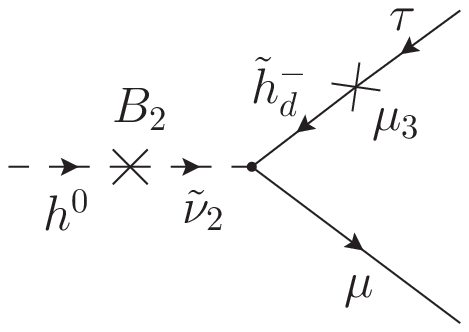}
\end{minipage}
\begin{minipage}[]{0.327\linewidth} 
\includegraphics[scale=0.51]{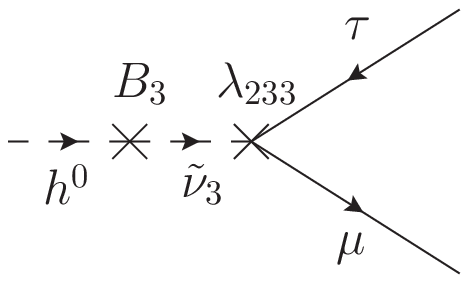}
\end{minipage}
\begin{minipage}[]{0.327\linewidth} 
\includegraphics[scale=0.51]{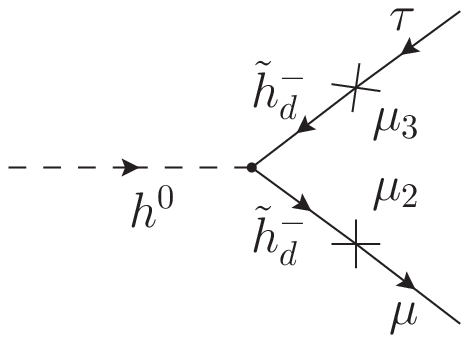}
\end{minipage}
\caption{Examples of $B_2 \mu_3$, $B_3 \lambda_{233}$ and $\mu_2 \mu_3$ contribution to $h^0 \rightarrow \mu^- \tau^+$ via the tree diagram.
%
%
}
\label{fig.1}
\end{figure}

Among all $B_i \mu_j$ combinations, only $B_2 \mu_3$ and $B_3 \mu_2$ 
are expected to give an important contribution to 
$h^0\rightarrow\mu^\mp \tau^\pm$ because they contribute already at 
tree level (see Fig. \ref{fig.1}, left panel). {In the case of $h^0\to \mu^- \tau^+$ the amplitude of tree level effective coupling can be approximated by 
\begin{align}
{\cal C}_L\!\cong
&\Big[\big.\!-g_{\scriptscriptstyle 2}\mu^*_2 B_3 m_2 M_W\frac{M_2\sin\!\beta+\mu_0\cos\!\beta}
  {\left(\mu_0 M_2\!-\!M^2_W\sin2\beta\right)^2\!\!M^2_s}\nonumber\\
&-\frac{y_{e_2}}{\sqrt{2}}
        \frac{\mu_3 B^*_2 M_2}{\left(\mu_0 M_2-M^2_W\sin2\beta\right) M^2_s}\big.\Big]\!
\left(\tan\beta\sin\alpha-\cos\alpha\right) \nonumber  \\    
%
{\cal C}_R\!\cong
&\Big[\big.\!-g_{\scriptscriptstyle 2}\mu_3 B^*_2 m_3 M_W\frac{M^*_2\sin\beta+\mu^*_0\cos\beta}{(\mu_0 M_2-M^2_W\sin2\beta)^2 M^2_s }\nonumber\\
&-\frac{y_{e_3}}{\sqrt{2}}\frac{\mu^*_2 B_3 M^*_2 }{(\mu_0 M_2-M^2_W\sin2\beta) M^2_s}\big.\Big]
(\tan\!\beta\sin\alpha-\cos\alpha) \,.
\end{align} 
$M_s$ denotes a generic real scalar mass eigenvalue.}
$B_3 \mu_2$ is particularly enhanced by tau Yukawa coupling thus becoming the largest among all $B_i \mu_j$'s. 
These two combinations get their most significant 
values under small $\mu_0$ and $M^2_s$ as can be seen from the formula. 

On the other hand, the values of $B_i$ and $B_i \mu_j$ are highly 
constrained separately by their loop contribution to neutrino mass 
matrix \cite{Rakshit}; a nonzero $\mu_j$ will induce tree level 
neutrino mass, hence it is very constrained. In the meantime, leptonic 
radiative decays like $\mu \rightarrow e\gamma$, etc. also give upper 
bounds on $B_i \mu_j$, e.g., $\left|B^*_2 \mu_3\right|$ and 
$\left|B_3 \mu^*_2\right| \lessapprox 10^{-4}\left|\mu_0\right|^3$  
\cite{ottoleptonic}.

{We require tree level neutrino mass and all the loop contributions 
from $B_i$ and $B_i \mu_j$ to neutrino mass matrix elements 
within 1 eV. The branching ratios we obtained are listed in Table \ref{table}.
With the neutrino mass constraint, we find that constraints from leptonic radiative decays are 
automatically satisfied. 
{For the cases of $B_3 \mu_2$ and $B_2 \mu_3$, though the one-loop contribution is roughly smaller than that from tree diagram, it can be still sizeable. A full discussion on the contribution from loop diagrams will be published in Ref.\cite{next}.}
Note that $B_2$ or $B_3$ alone can give contributions to 
$h^0\rightarrow\mu^\mp \tau^\pm$ as well. 
Such contributions are included in the $B_i \mu_j$ result
(and $B_i \lambda$, $B_i A^\lambda$ thereafter).
\subsection{Contributions from  $B_i\,\lambda$ combinations}

Among all $B_i \lambda$'s, it is obvious that 
$B_1 \lambda_{123}$, $B_1 \lambda_{132}$, 
$B_2 \lambda_{232}$ and $B_3 \lambda_{233}$ are the most significant
because they can already contribute to the amplitudes at the tree level (Fig. \ref{fig.1}, middle panel). This
tree level contribution, {in the case of $h^0\to \mu^- \tau^+$, can be approximated by 
\begin{align} 
{\cal C}_L&\cong 
\frac{\lambda_{3j2}}{\sqrt{2}}
       \frac{B^*_j}{M^2_s} \left(\tan\beta\sin\alpha-\cos\alpha\right)\nonumber\\
{\cal C}_R&\cong 
\frac{\lambda^*_{2j3}}{\sqrt{2}} 
        \frac{B_j}{M^2_s}(\tan\!\beta\sin\alpha-\cos\alpha) \,,
\end{align} }
where $\alpha$ is the mixing angle between the two CP-even neutral Higgs.

The value of $B_i$ is constrained by $B_i B_j$ 
neutrino mass loops \cite{Rakshit}, while that of $\lambda$ is constrained
by charged current experiments \cite{Barbier}. 
Generally speaking, increasing the supersymmetry breaking slepton mass and
the gaugino mass leads to heavier charged slepton, sneutrino and neutralino, and 
hence makes upper bounds for $B_i$ and $\lambda$ raise. Due to the strong 
dependence upon $B_i$ and $\lambda$ by tree level amplitude,  
this approach is still favorable even though heavy sneutrinos tend to suppress 
the branching ratio.
$\mu_0$ should not be too small, in order to keep 
 the product of $B_i$ and $\lambda$ below the bounds from leptonic 
radiative decays, i.e., $\left|B^*_1 \lambda_{132}\right|$, 
$\left|B_1 \lambda^*_{123}\right|$, $\left|B^*_2 \lambda_{232}\right|$ 
and $\left|B_3 \lambda^*_{233}\right| \lessapprox 1.4\times10^{-3}
\left|\mu_0\right|^2$ 
\cite{ottoleptonic}. Even under the stringent neutrino mass $\lesssim$ 1 eV 
constraint, we find that the four combinations give
  sizeable branching ratios of about $10^{-5}$ (Fig. \ref{fig.3}, upper panel), which {may be large enough to be probed at the LHC.}

\begin{table}[ht]
\caption{Contributions from interesting combinations of RPV parameters to 
$Br(h^0\rightarrow\mu^\mp \tau^\pm)$}\vspace{-18pt}
\label{table}
\begin{center}
		\begin{tabular}{c@{\hspace{20pt}}c@{\hspace{20pt}}c}
		\multicolumn{2}{l}{} \\
		\hline\hline\\[-10pt]
				RPV Parameter  & $Br$ with Neutrino\\[0pt]
		Combinations  & Mass $\lesssim$ 1 eV Constraint\\
		\hline \\[-10pt]
		           $B_2 \,\mu_3\hspace*{8pt} $  & $1\times10^{-15}$ \\
		           $B_3 \,\mu_2\hspace*{8pt} $  & $1\times10^{-13}$ \\
		           		\hline\\[-10pt]
               $B_1 \,\lambda_{123}$  & $1\times10^{-5\hspace{4pt}}$ \\
               $B_1 \,\lambda_{132}$  & $3\times10^{-5\hspace{4pt}}$ \\
               $B_2 \,\lambda_{232}$  & $3\times10^{-5\hspace{4pt}}$ \\
               $B_3 \,\lambda_{233}$  & $3\times10^{-5\hspace{4pt}}$ \\
               		\hline\\[-10pt]
               $\mu_2\,\mu_3\hspace*{8pt}$  & $2\times10^{-18}$\\ 
               		\hline\\[-10pt]
               $B_1 \,A^\lambda_{123}$  & $5\times10^{-11}$\\
           $B_1 \,A^\lambda_{132}$  &$5\times10^{-11}$\\
           $B_2 \,A^\lambda_{232}$  &$5\times10^{-11}$\\
           $B_3 \,A^\lambda_{233}$  &$5\times10^{-11}$\\[2pt] 
		\hline
		\end{tabular}\\
		\end{center}
\end{table} 
\subsection{Contributions from  $\mu_i\,\mu_j$ combinations}
{As to $\mu_i \mu_j$ combinations, up to one-loop level only $\mu_2\,\mu_3$ contributes to $h^0\rightarrow\mu^\mp \tau^\pm$ (Fig. \ref{fig.1}, right panel). In the case of $h^0\to \mu^- \tau^+$ the amplitude of tree-level contribution can be approximated as
\begin{align}
{\cal C}_L\cong
&-g_{\scriptscriptstyle 2}\mu^*_2 \mu_3 m_2 M_W \cos\beta \frac{M^2_2+2M^2_W \cos^2\!\beta}
       {\left(\mu_0 M_2-M^2_W\sin2\beta\right)^3} \cos\alpha  \nonumber\\
&-g_{\scriptscriptstyle 2}\mu^*_2 \mu_3 m_2 M_2 M_W\frac{M_2\sin\!\beta+\mu_0\cos\!\beta}
  {\left(\mu_0 M_2-M^2_W\sin2\beta\right)^3}\sin\alpha   \nonumber\\
%
{\cal C}_R\cong&   
-g_{\scriptscriptstyle 2}\mu^*_2 \mu_3 m_3 M_W\cos\!\beta \frac{M^2_2+2M^2_W\cos^2\beta}{(\mu_0 M_2-M^2_W\sin2\beta)^3}
        \cos\alpha \nonumber\\
&-g_{\scriptscriptstyle 2}\mu^*_2 \mu_3 m_3 M^*_2 M_W\frac{M^*_2\sin\beta+\mu^*_0\cos\beta}{(\mu_0 M_2-M^2_W\sin2\beta)^3}
        \sin\alpha \,.
\end{align}
 With a nonzero $\mu_i$\,, one of the neutrinos gets a tree level mass. However, leptonic radiative
decays set a more stringent bound on $\mu_2\,\mu_3$ than neutrino mass does \cite{ottomu}.
Interestingly enough, though the  $\mu_2 \mu_3$ combination contributes to the decay in tree level, contribution from the loop diagram (Fig. \ref{fig.2}, left panel) is generally more important due to the small neutrino mass in the loop.
Unfortunately, $\mu_2 \mu_3$ could only give a negligible branching ratio because of the stringent bound on $\mu_2 \mu_3$.} 
 
\subsection{Contributions from  $B_i\,A^\lambda$ combinations}
The contributions from $B_i\,A^\lambda$ combinations are quite interesting because they will be like the
first experimental signature of the RPV $A$-parameters. However,
$A^\lambda$ only plays its role in a single loop diagram (Fig. \ref{fig.2}, right panel) via the neutral 
scalar-charged scalar-charged scalar ($h^0\phi^+\phi^-$) coupling. 
It is expected to give important contributions for low charged scalar mass
to avoid strong suppression. $A^\lambda$ is not constrained by the radiative decays \cite{ottoleptonic}
and can consequently take any value \footnote{Recall that under the parametrization adopted here, the $\hat{L}_i$ superfields all have zero VEV. Hence, $A^\lambda$ as defined
does not contribute, for example, to $b \to s\gamma$ at one-loop level.}. But $B_i$ is still limited by 
 neutrino mass loops as before. 

\begin{figure}[ht]
\begin{minipage}[]{0.495\linewidth} 
\includegraphics[scale=1.0]{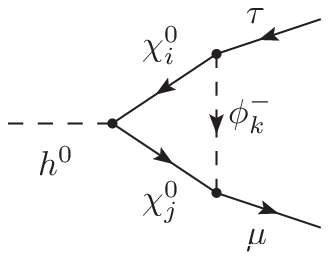}
\end{minipage}
\begin{minipage}[]{0.495\linewidth} 
\includegraphics[scale=1.0]{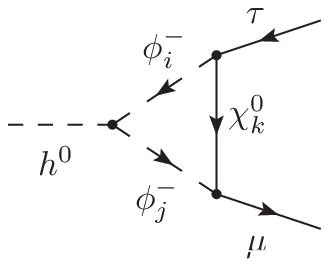}
\end{minipage}
\caption{
(Left Panel) The one-loop diagram which can give a more significant contribution in the case of $\mu_2\mu_3$ combination than the tree diagram does.
(Right Panel) The only one-loop diagram in which $A^\lambda$ has its stage. $A^\lambda$ participates in via the $h^0\phi^-_i\phi^-_j$ coupling.}
\label{fig.2}
\end{figure}
Under the condition of $A^\lambda$=2500 GeV, the branching ratios from $B_i A^\lambda$ can 
reach a maximum order of $10^{-11}$. 
Since decay rate is proportional to amplitude 
square and hence $A^\lambda$ square, it is easy to see how branching 
ratio changees as $A^\lambda$ increases.
{As an example, we illustrate in Fig. \ref{fig.3} (lower panel) the branching ratio from $B_2 A^\lambda_{232}$ contribution for $A^\lambda_{232}=$2500 GeV and 2500 TeV. In the extreme case of
$A^\lambda_{232}=$2500 TeV, the branching ratio is 6 orders of magnitude larger than that in the case of $A^\lambda_{232}=$2500 GeV and achieves the
order of $10^{-5}$. }
\section{Summary}
There are different scenarios to achieve LFV Higgs decay in the supersymmetric standard model depending on how neutrino masses are implemented. We have shown that even with RPV parameters only, significant contributions to 
$h^0\to \mu^\mp\tau^\pm$ are possible. In our analysis, $B_i\, \lambda$'s 
undoubtedly provide the largest branching ratios. 
Even with stringent neutrino mass constraint $m_\nu\lesssim\text{1 eV}$, 
several combinations of $B_i\, \lambda$ can still give branching 
ratios beyond $10^{-5}$ which should not be overlooked in future 
collider experiments. 

At the end, we would like to encourage our experimentalist 
colleagues at CMS and ATLAS to investigate those flavor violating 
Higgs decays because they may give some complementary information 
about lepton flavor violating couplings, which are also relevant for 
neutrinos physics.
A typical cross-section of MSSM 125 GeV Higgs
at 8 TeV energy is of the order 10 pb. With a luminosity of the order $10$ fb$^{-1}$, we estimated for the
$Br(h^0\to \mu^\mp \tau^\pm)$ being of the order $10^{-5}$ that
it would lead to 
several raw $\mu^\mp \tau^\pm$ events with almost no SM background.
Our estimate is likely to be on the optimistic side when detector properties are fully taken into consideration.
Some complete experimental analyses with realistic cuts may be needed to improve the situation.
Note however that after we finished our work a preprint \cite{simulation} on the relevant branching ratio reach
of the 8 TeV LHC appears, claiming a quite disappointing number.
The case for the 14 TeV running may provide a better chance to probe the lepton flavor violating couplings.
If we allow more free parameters or a larger parameter space during our
analysis, the branching ratios can become larger. The signal should certainly be a focus of any dedicated Higgs machine.

\begin{figure}[ht]
\begin{minipage}[]{0.95\linewidth} 
\onefigure[scale=1.0]{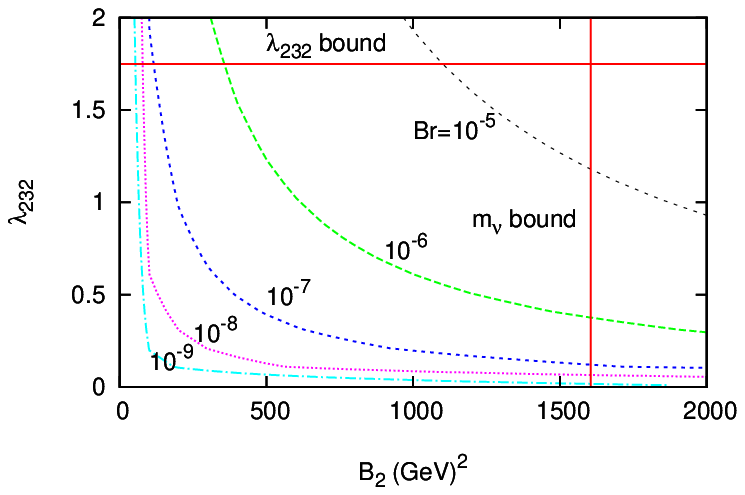}
\end{minipage}\vspace{-7pt}
\begin{minipage}[]{0.95\linewidth} 
\onefigure[scale=1.0]{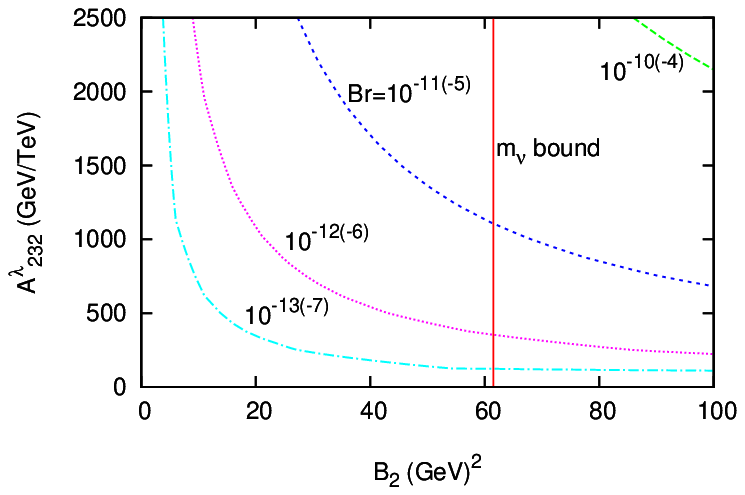}
\end{minipage}
\caption{Solid vertical line ($m_\nu$ bound) comes from demanding 22-element of neutrino mass matrix $<$1 eV.
(Upper Panel) Branching ratio from $B_2\,\lambda_{232}$, with $M_2=2500$ GeV, $\tilde{m}^2_{L_{ii}}=\tilde{m}^2_{E_{ii}}=$(2500 GeV)$^2$, $\mu_0=1900$ GeV$=A_u=-A_d$\,, $\tan\beta=60$. $M_A\cong$ 200 to 205 GeV in the permitted region.
(Lower Panel) Branching ratio from $B_2\,A^\lambda_{232}$, with $M_2=$2500 GeV, $\tilde{m}^2_{L_{ii}}=\tilde{m}^2_{E_{ii}}$=(500 GeV)$^2$, $\mu_0=$1800 GeV$=A_u=-A_d$\,, $\tan\beta=60$. $M_A\cong$200 GeV.
$A^\lambda_{232}$ ranges from 0 to 2500 GeV(TeV). }
\label{fig.3}
\end{figure} 

Below is some estimate about what to expect in a Higgs factory.
There is mainly 2 processes which can contribute to the Higgs production, i.e.,
$e^+ e^-\to ZH$ and $e^+ e^-\to \nu_e \bar{\nu}_e H$. For $e^+ e^-\to ZH$, the cross-section of a 125 GeV SM Higgs boson is roughly 200 fb near the threshold. With a luminosity of 500 fb$^{-1}$ and $\sin(\beta-\alpha)\cong 1$, we may have several raw events of $h^0\to \mu^\mp \tau^\pm$ for a branching ratio of the order $10^{-5}$.
At a higher energy, taking 3 TeV as an example, the process which provides a large cross-section is $e^+ e^-\to \nu_e \bar{\nu}_e H$ via $WW$ fusion \cite{higgsfactory}. In this case the cross-section is about 500 fb. With a luminosity of 1000 fb$^{-1}$ and $Br(h^0\to \mu^\mp \tau^\pm)\gtrsim 10^{-5}$, we may have several tens of raw events.
\acknowledgments
Y.C. and O.K. are partially supported by research grant NSC 99-
2112-M-008-003-MY3 of the National Science Council of Taiwan. We thank Amy Brainer for her help in improving the English of the text.  
\end{document}